\documentstyle[prd,aps,amsmath,latexsym,amssymb,floats]{revtex}
\def \slas{\kern -6.2pt /}
\def \sla{\kern -5.4pt /}
\def \sl{\kern -4.0pt /}
\def \Cslas{\kern -6.8pt /}
\def \Dslas{\kern -7.4pt /}
\def \slass{\kern -7.4pt /}

\def \ii{{\mathrm{i}}}
\def \d{{\mathrm{d}}}

\def \pd{\partial}
\def \e{{\mathrm{e}}}

\def \lcx{\tilde{x}}
\def \tl#1{\overset{\kern 2pt\circ}{#1}}
\def \tll#1{\overset{\kern -1pt\circ}{#1}}
\def \TL#1{\overset{\kern -28pt \circ}{#1}}
\def \TLL#1{\overset{\kern -7pt \circ}{#1}}

\def \LD#1{\overset{\,\leftarrow}{#1}}
\def \RD#1{\overset{\,\,\rightarrow}{#1}}

\newcommand{\gvh}{\hat g_\perp^{(v)}}
\newcommand{\gah}{\hat g_\perp^{(a)}}
\newcommand{\gvn}{g_{\perp n}^{(v)}}

\newcommand{\hs}{h_\parallel^{(s)}}
\newcommand{\hsh}{\hat h_\parallel^{(s)}}

\newcommand{\htt}{h_\parallel^{(t)}}
\newcommand{\htth}{\hat h_\parallel^{(t)}}

\newcommand{\httn}{h_{\parallel n}^{(t)}}

\begin{document}

\title{Wandzura-Wilczek-type relations of $\rho$-meson wave functions}
\author{Markus Lazar\thanks{E-mail: lazar@itp.uni-leipzig.de}}
\address{Centre for Theoretical Studies
         and Institute of Theoretical Physics,\\
        Leipzig University, 
        Augustusplatz~10, D-04109~Leipzig, Germany} 
\date{\today}    
        
\maketitle
\hspace*{1cm}
\begin{abstract}
We give the geometric Wandzura-Wilczek-type relations between the meson wave functions
of dynamical twist by means of the meson wave functions of geometric twist.
We discuss the difference between geometric and dynamical Wandzura-Wilczek-type
relations.
Additionally, the interrelations between the different twist notations of 
meson wave functions are discussed. 
 
\end{abstract}

\section{Introduction}
In a recent paper~\cite{L00}, I have introduced the (two-particle) $\rho$-{\em meson wave 
functions} which are related to nonlocal LC-operators of different
geometric twist~\cite{glr99,gl00a}. 
This twist decomposition of nonlocal operators is based on the notion of 
{\it geometric} twist = mass dimension -- (Lorentz) spin, $\tau=d-j$, 
originally introduced by Gross and Treiman~\cite{Gross}. 
I found eight meson wave functions of geometric twist.

On the other hand, Ball {\it et al.} used the notion of {\it dynamical twist} $t$,
which was introduced by Jaffe and Ji~\cite{Jaf92} by counting powers $Q^{2-t}$,
for the classification of (two-particle) vector meson wave functions and 
vector meson 
distribution amplitudes, respectively~\cite{ball96,ball98,ball99}. 
They classified the meson wave functions 
which correspond to the independent tensor structure of the matrix 
elements of bilocal quark-antiquark operators.
One key ingredient in their approach was the use of QCD equations of motion
in order to obtain dynamical Wandzura-Wilczek-type relations for wave 
functions that are not dynamically independent.
The pion wave functions were investigated in similar way~\cite{braun89,braun90,ball99b}.
A systematic study of Wandzura-Wilczek-type relations for the forward case 
is given in Ref.~\cite{gl00d}.

In the framework of geometric twist, it is possible to investigate 
the interrelations and the mismatch between the different twist 
definitions of meson wave functions as a kind of group theoretical relations. 
By means of this relations, we are able to obtain geometric
Wandzura-Wilczek-type relations between the the dynamical twist function
which differ from the dynamical Wandzura-Wilczek-type relations. 

\section{The dynamical twist $\rho$-meson wave functions}
Usually, the meson light-cone distributions are defined as the 
vacuum-to-meson matrix elements of quark-antiquark nonlocal gauge invariant 
operators on the light-cone.
In this way,
Ball~{\it et al.}~\cite{ball96,ball98,ball99} found eight independent two-particle distributions.
The classification of these dynamical twist $\rho$-meson wave functions with respect to spin, dynamical twist, chirality and the relations to geometrical twist (see section~III)
are summarized in Tab.~\ref{tab:2}. 
\begin{table}
\begin{center}
\renewcommand{\arraystretch}{1.3}
\begin{tabular}{|c|ccc|}
\hline
Twist $t$   & 2 & 3 & 4 \\
    & $O(1)$  & $O(1/Q)$& $O(1/Q^{2})$ \\ \hline
$e_{\parallel}$ & $\hat\phi_{\parallel}\equiv\hat\Phi^{(2)}$ & 
$\underline{\htth}=\hat\Phi^{(3)}+F_t(\hat\Phi^{(2)},\hat\Phi^{(3)})$, $\underline{\hsh}\equiv\hat\Upsilon^{(3)}$& 
$\hat g_{3}=\hat\Phi^{(4)}+F_g(\hat\Phi^{(2)},\hat\Phi^{(3)},\hat\Phi^{(4)})$ \\
$e_{\perp}$ & $\underline{\hat\phi_{\perp}}\equiv\hat\Psi^{(2)}$ & 
$\hat g_{\perp}^{(v)}=\hat\Psi^{(3)}+F_v(\hat\Psi^{(2)},\hat\Psi^{(3)})$,
$\hat g_{\perp}^{(a)}\equiv\hat\Xi^{(3)}$ & 
$\underline{\hat h_{3}}=\hat\Psi^{(4)}+F_h(\hat\Psi^{(2)},\hat\Psi^{(3)},\hat\Psi^{(4)})$
\\[2pt] \hline
\end{tabular}
\renewcommand{\arraystretch}{1}
\end{center}
\caption{Spin, dynamical twist and chiral classification of the $\rho$-meson
distribution 
amplitudes.}
\label{tab:2}
\end{table}%
One distribution amplitude was obtained 
for longitudinally ($e_{\parallel}$) and transversely ($e_{\perp}$)
polarized $\rho$-mesons of twist-2 and twist-4, respectively.
Whereas, the number of  twist-3
distribution amplitudes is doubled for each polarization.
The higher twist distribution amplitudes
contribute to a hard exclusive amplitude with
additional powers of $1/Q$ compared to the leading twist-2 ones.
The underlined distribution amplitudes 
are chiral-odd, the others chiral-even.

By using the Lorentz decomposition of the matrix elements of relevant operators in analogy to nucleon structure functions, the explicit definitions of the chiral-even $\rho$-distributions 
are\footnote{Here I am using the symmetric wave function $\hat\phi(\xi)$ 
(see also~\cite{chern84}) instead of
$\phi(u)$ because the corresponding integral relations will be easier.}:
\begin{align}
\langle 0|\bar u(\lcx) \gamma_{\alpha} U(\lcx,-\lcx)d(-\lcx)|\rho(P,\lambda)\rangle 
&= f_{\rho} m_{\rho} \left[ p_{\alpha}
\frac{e^{(\lambda)}\lcx}{\lcx P}
\int_{-1}^{1} \!\d \xi\, \e^{\ii \xi(\lcx P)} \hat\phi_{\parallel}(\xi, \mu^{2})  
+ e^{(\lambda)}_{\perp \alpha}
\int_{-1}^{1} \!\d \xi\, \e^{\ii \xi(\lcx P)} \hat g_{\perp}^{(v)}(\xi, \mu^{2}) \right.
\nonumber \\
&\qquad\left.-\frac{1}{2}\lcx_{\alpha}
\frac{e^{(\lambda)}\lcx }{(\lcx P)^{2}} m_{\rho}^{2}
\int_{-1}^{1} \!\d \xi\, \e^{\ii \xi (\lcx P)} \hat g_{3}(\xi, \mu^{2}) \right]
\label{eq:vda}
\end{align}
and 
\begin{eqnarray}
\langle 0|\bar u(\lcx) \gamma_{\alpha} \gamma_{5}U(\lcx,-\lcx)
d(-\lcx)|\rho(P,\lambda)\rangle = 
\frac{1}{2}\left(f_{\rho} - f_{\rho}^{\rm T}
\frac{m_{u} + m_{d}}{m_{\rho}}\right)
m_{\rho} \epsilon_{\alpha}^{\phantom{\alpha}\beta \mu \nu}
e^{(\lambda)}_{\perp \beta} p_{\mu} \lcx_{\nu}
\int_{-1}^{1} \!\d \xi\, \e^{\ii \xi(\lcx P)}\hat g^{(a)}_{\perp}(\xi, \mu^{2}),
\label{eq:avda}
\end{eqnarray}
while the chiral-odd distributions are defined as 
\begin{align}
\langle 0|\bar u(\lcx) \sigma_{\alpha \beta} U(\lcx,-\lcx)d(-\lcx)|\rho(P,\lambda)\rangle 
&= \ii f_{\rho}^{\rm T} \left[ ( e^{(\lambda)}_{\perp \alpha}p_\beta -
e^{(\lambda)}_{\perp \beta}p_\alpha )
\int_{-1}^{1} \!\d \xi\, \e^{\ii \xi(\lcx P)} \hat\phi_{\perp}(\xi,\mu^2) \right.
\nonumber \\
&\qquad + (p_\alpha \lcx_\beta - p_\beta \lcx_\alpha )
\frac{e^{(\lambda)}\lcx}{(\lcx P)^{2}}
m_{\rho}^{2} 
\int_{-1}^{1} \!\d \xi\, \e^{\ii \xi(\lcx P)} \htth (\xi, \mu^{2}) 
\nonumber \\
& \qquad \left.+ \frac{1}{2}
(e^{(\lambda)}_{\perp \alpha} \lcx_\beta -e^{(\lambda)}_{\perp \beta} \lcx_\alpha) 
\frac{m_{\rho}^{2}}{\lcx P} 
\int_{-1}^{1} \!\d \xi\, \e^{\ii \xi(\lcx P)}\hat h_{3}(\xi, \mu^{2}) \right]
\label{eq:tda}
\end{align}
and
\begin{equation}
\langle 0|\bar u(\lcx) U(\lcx,-\lcx)
d(-\lcx)|\rho(P,\lambda)\rangle
= -\ii \left(f_{\rho}^{\rm T} - f_{\rho}\frac{m_{u} + m_{d}}{m_{\rho}}
\right)(e^{(\lambda)}\lcx) m_{\rho}^{2}
\int_{-1}^{1} \!\d \xi\, \e^{\ii \xi(\lcx P)}\hat \hs(\xi, \mu^{2}).
\label{eq:sda}
\end{equation}
In the above definitions we used the two light-cone vectors
\begin{align}
\label{Osc}
\lcx_\alpha = x_\alpha -\frac{P_\alpha}{m_\rho^2} \Big(
(xP) - \sqrt{(xP)^2 - x^2 m_\rho^2}
\Big), \qquad
p_\alpha=P_\alpha-\frac{1}{2} \lcx_\alpha\frac{m_\rho^2}{\lcx P},
\end{align}
with $p^2=0$, $\lcx^2=0$ and $p\cdot\lcx=P\cdot\lcx$.
Here $P_\alpha$ is the $\rho$-meson momentum vector, so that
$P^2=m_\rho^2$, $e^{(\lambda)}\cdot e^{(\lambda)}=-1$, $P\cdot e^{(\lambda)}=0$, 
$m_\rho$ denotes the $\rho$-meson mass.
The polarization vector $e^{(\lambda)}_\alpha$ 
can be decomposed into projections onto the two light-like vectors and the 
orthogonal plane:
\begin{align}
e^{(\lambda)}_\alpha=p_\alpha \frac{e^{(\lambda)}\lcx}{\lcx P}-
\frac{1}{2} \lcx_\alpha \frac{e^{(\lambda)}\lcx}{(\lcx P)^2} m_\rho^2
+e^{(\lambda)}_{\bot\alpha}.
\end{align}

The distribution amplitudes are dimensionless functions of $\xi$ and 
describe
the probability amplitudes to find the $\rho$ in a state with minimal
number of constituents (quark and antiquark) which carry
momentum fractions $\xi$.
The nonlocal operators are renormalized at scale $\mu$, so that
the distribution amplitudes depend on $\mu$ as well. This
dependence can be calculated in perturbative QCD (from now on we suppress
$\mu^2$).

The vector and tensor  decay constants $f_\rho$ and $f_\rho^{\rm T}$ are defined 
as usually as
\begin{eqnarray}
\langle 0|\bar u(0) \gamma_{\alpha}
d(0)|\rho(P,\lambda)\rangle & = & f_{\rho}m_{\rho}
e^{(\lambda)}_{\alpha},
\label{eq:fr}\\
\langle 0|\bar u(0) \sigma_{\alpha \beta} 
d(0)|\rho(P,\lambda)\rangle &=& \ii f_{\rho}^{\rm T}
(e_{\alpha}^{(\lambda)}P_{\beta} - e_{\beta}^{(\lambda)}P_{\alpha}).
\label{eq:frp}
\end{eqnarray}
All eight distributions $\hat\phi=\{\hat\phi_\parallel,\hat \phi_\perp,
\hat g_\perp^{(v)},\hat g_\perp^{(a)},\htth,\hsh,\hat h_3,\hat g_3\}$ 
are normalized as
\begin{equation}
\int_{-1}^1\!\d \xi\, \hat\phi(\xi) =1.
\label{eq:norm}
\end{equation}
The wave function moments are given as
\begin{align}
\label{moment}
\phi_n=\int_{-1}^1\!\d \xi\, \xi^n \hat\phi(\xi)
\equiv\int_{0}^1\!\d u\, \xi^n \phi(u),\qquad \xi=2u-1
\end{align}
with both types of meson wave functions  
\begin{align}
\hat\phi(\xi)=\frac{1}{2}\,\phi\left(\frac{\xi+1}{2}\right),\qquad
\phi(u)=2\hat\phi(2u-1).
\end{align}
Eq.~(\ref{moment}) means, having information about the wave function moments, one can 
reconstruct the wave function itself.

\section{The geometric twist $\rho$-meson wave functions and relations to dynamical twist meson wave functions}
Reproducing results of~\cite{L00}, we define the $\rho$-meson wave functions of
geometric twist. By means of operators of geometric twist~\cite{L00}, 
the chiral-even $\rho$-distributions 
with definite (geometric) twist are:
\begin{align}
\label{O_v}
&\langle 0|\bar u(\lcx) \gamma_{\alpha} U(\lcx,-\lcx)d(-\lcx)|\rho(P,\lambda)\rangle
=f_\rho m_\rho\bigg[ p_\alpha\frac{e^{(\lambda)}\lcx}{\lcx P}\int_{-1}^1\d \xi\, \hat\Phi^{(2)}(\xi) e_0(\ii\zeta\xi)\\
&\qquad+e^{(\lambda)}_{\perp\alpha}\int_{-1}^1\!\!\d \xi
\Big\{\hat\Phi^{(2)}(\xi)e_1(\ii\zeta \xi)+\hat\Phi^{(3)}(\xi)
[e_0(\ii\zeta \xi)-e_1(\ii\zeta \xi)]\Big\}
\nonumber\\
&\qquad
-\frac{1}{2}\lcx_\alpha\frac{e^{(\lambda)}\lcx}{(\lcx P)^2} m_\rho^2
\int_{-1}^1\!\!\d \xi
\Big\{\hat\Phi^{(4)}(\xi)\Big[e_0(\ii\zeta \xi)-3e_1(\ii\zeta \xi)
+2\int_0^1\!\!\d t\,  e_1(\ii\zeta \xi t)\Big]
-\hat\Phi^{(2)}(\xi)\Big[e_1(\ii\zeta \xi)
-2\int_0^1\!\!\d t\, e_1(\ii\zeta \xi t)\Big]
\nonumber\\
&\qquad\qquad\qquad\qquad\qquad\qquad
+4\hat\Phi^{(3)}\Big[e_1(\ii\zeta \xi)
-\int_{0}^1\!\!\d t\, e_1(\ii\zeta \xi t)\Big]\Big\}\bigg]\nonumber
\end{align}
and
\begin{align}
\label{O_5v}
&\langle 0|\bar u(\lcx) \gamma_{\alpha}\gamma_5 U(\lcx,-\lcx)d(-\lcx)|\rho(P,\lambda)\rangle
=
\frac{1}{2}\bigg(f_\rho -f_\rho^{\rm T}\frac{m_u+m_d}{m_\rho}\bigg)m_\rho
\epsilon_\alpha^{\ \,\beta\mu\nu}
e^{(\lambda)}_{\perp\beta} p_\mu\lcx_\nu
\int_{-1}^1\d \xi\, \hat\Xi^{(3)}(\xi) e_0(\ii\zeta \xi),
\end{align}
while the chiral-odd distributions are defined as
\begin{align}
\label{M_ten}
&\langle 0|\bar u(\lcx) \sigma_{\alpha \beta} U(\lcx,-\lcx)d(-\lcx)|\rho(P,\lambda)\rangle
=
\ii f_\rho^{\rm T}\bigg[
2e^{(\lambda)}_{\bot[\alpha}p_{\beta]}
\int_{-1}^1\!\!\d \xi \hat\Psi^{(2)}(\xi)e_0(\ii\zeta \xi)
\\
&\qquad
+2\lcx_{[\alpha}p_{\beta]}\frac{e^{(\lambda)}\lcx}{(\lcx P)^2}m_\rho^2
\int_{-1}^1\!\!\d \xi \Big\{
 2\hat\Psi^{(2)}(\xi)e_2(\ii\zeta \xi)
-\hat\Psi^{(3)}(\xi)\big[e_0(\ii\zeta \xi)+2e_2(\ii\zeta \xi)\big]
\Big\}
\nonumber\\
&\qquad
-\lcx_{[\alpha}e^{(\lambda)}_{\beta]\perp}\frac{m_\rho^2}{\lcx P}
\int_{-1}^1\!\!\d \xi \Big\{
2 \hat\Psi^{(2)}(\xi)\big[e_1(\ii\zeta \xi)+e_2(\ii\zeta \xi)\big]
- \hat\Psi^{(3)}(\xi)\big[1+2e_2(\ii\zeta \xi)\big]
+ \hat\Psi^{(4)}(\xi)\big[1+e_0(\ii\zeta \xi)-2e_1(\ii\zeta \xi)\big]
\Big\}\bigg] \nonumber
\end{align}
and
\begin{align}
\label{O_s}
\langle 0|\bar{u}(\lcx)
U(\lcx,-\lcx)d(-\lcx)|\rho(P,\lambda)\rangle
&=-\ii\bigg( f^{\rm T}_\rho-f_\rho \frac{m_u+m_d}{m_\rho}\bigg)
\big(e^{(\lambda)}\lcx\big)m_\rho^2
\int_{-1}^1\d \xi\, \hat\Upsilon^{(3)}(\xi)e_0(\ii\zeta \xi).
\end{align}
Here we used the following ``truncated exponentials'' (with $\zeta=(\lcx P)$)
\begin{align}
e_0(\ii\zeta \xi)=\e^{\ii\zeta \xi},\qquad 
e_1(\ii\zeta \xi)=\int_0^1\!\!\d t \,\e^{\ii\zeta \xi t}
=\frac{\e^{\ii\zeta \xi}-1}{\ii\zeta \xi},
\quad\cdots\quad,
e_{n+1}(\ii\zeta \xi)
=\frac{(-1)^{n}}{n!}\int_0^1\!\!\d t \, t ^n\,\e^{\ii\zeta \xi t}.
\end{align}

Comparing these expressions~(\ref{O_v})--(\ref{O_s})
with the meson wave functions of dynamical twist
(\ref{eq:vda}) -- (\ref{eq:sda}),
we observe that it is necessary to re-express the truncated exponentials
and perform appropriate variable transformations. After such manipulations
we obtain the following relations, which allow to reveal the interrelations
between the different twist definitions of meson wave functions:
\begin{align}  
\label{rel-phi1}
\hat\phi_\|(\xi)&\equiv\hat\Phi^{(2)}(\xi),\\
\label{rel-gv}
\gvh(\xi)&=\hat\Phi^{(3)}(\xi) + \int_\xi^{{\rm sign}(\xi)} \frac{\d \tau}{\tau}
\Big(\hat\Phi^{(2)}-\hat\Phi^{(3)}\Big)\left(\tau\right),\\
\label{rel-g3}
\hat g_3(\xi)&=\hat\Phi^{(4)}(\xi) - \int_\xi^{{\rm sign}(\xi)} \frac{\d\tau}{\tau}\Big\{
\Big(\hat\Phi^{(2)}-4\hat\Phi^{(3)}+3\hat\Phi^{(4)}\Big)\left(\tau\right)
+ 2 \ln \Big(\frac{\xi}{\tau}\Big)
\Big(\hat\Phi^{(2)}-2\hat\Phi^{(3)}+\hat\Phi^{(4)}\Big)\left(\tau\right)
\Big\},\\
\label{rel-ga}
\hat g_\perp^{(a)}(\xi)&\equiv\hat\Xi^{(3)}(\xi),\\
\label{rel-phi2}
\hat\phi_\perp(\xi)&\equiv\hat\Psi^{(2)}(\xi),\\
\label{rel-ht}
\hat h^{(t)}_\|(\xi)&=\hat\Psi^{(3)}(\xi) + 2\xi \int_\xi^{{\rm sign}(\xi)}\frac{\d \tau}{\tau^2}
\Big(\hat\Psi^{(2)}-\hat\Psi^{(3)}\Big)\left(\tau\right),\\
\label{rel-h3}
\hat h_3(\xi)&=\hat\Psi^{(4)}(\xi)+ \int_\xi^{{\rm sign}(\xi)} \frac{\d \tau}{\tau}\Big\{
2\Big(\hat\Psi^{(2)}-\hat\Psi^{(4)}\Big)\left(\tau\right)
-2\frac{\xi}{\tau}\Big(\hat\Psi^{(2)}-\hat\Psi^{(3)}\Big)\left(\tau\right)
-\delta\Big(\frac{\xi}{\tau}\Big)\Big(\hat\Psi^{(3)}-\hat\Psi^{(4)}\Big)\left(\tau\right)
\Big\},\\
\hat h_\|^{(s)}(\xi)&\equiv\hat\Upsilon^{(3)}(\xi).
\end{align}
Obviously, both decompositions coincide in the leading terms, e.g., 
the meson wave functions $\hat\phi_\|(\xi)$ and $\hat\phi_\perp(\xi)$ are genuine
geometric twist-2 and $\gah(\xi)$ and $\hsh(\xi)$ are genuine 
geometric twist-3 functions.
But the different twist definitions of meson wave functions differ at 
higher twist. For instance, the meson wave functions 
$\gvh(\xi)$ and $\hat h^{(t)}_\|(\xi)$ with dynamical twist $t=3$ contain 
contributions with geometrical twist $\tau = 2$ and $3$. Additionally,
dynamical twist $t=4$ meson wave functions  
$\hat g_3(\xi)$ and $\hat h_3(\xi)$ contain 
contributions with geometrical twist $\tau = 2$, $3$ as well as $4$.

The conventional wave functions can be written in
terms of the new $\rho$-meson wave functions.
The nontrivial relations are:
\begin{align}
\label{rel-Phi3}
\hat\Phi^{(3)}(\xi)&=\gvh(\xi) +\frac{1}{\xi} \int_\xi^{{\rm sign}(\xi)} \d \tau
\big(\gvh-\hat\phi_\|\big)\left(\tau\right),\\
\hat\Phi^{(4)}(\xi)&=\hat g_3(\xi) 
+\frac{1}{\xi^2} \int_\xi^{{\rm sign}(\xi)} \d \tau\, \tau
\big(3\hat g_3-4\gvh+\hat\phi_\|\big)\left(\tau\right)
+\frac{1}{\xi^2} \int_\xi^{{\rm sign}(\xi)} \d \tau\, \tau \Big(1-\frac{\xi}{\tau}\Big)
\big(\hat g_3-4\gvh+3\hat\phi_\|\big)\left(\tau\right),\\
\label{rel-Psi3}
\hat\Psi^{(3)}(\xi)&=\hat h_\|^{(t)}(\xi) + \frac{2}{\xi}\int_\xi^{{\rm sign}(\xi)} \d \tau
\big(\hat h_\|^{(t)}-\hat\phi_\perp\big)\left(\tau\right), \\
\hat\Psi^{(4)}(\xi)&=\hat h_3(\xi) 
+\frac{2}{\xi^2} \int_\xi^{{\rm sign}(\xi)} \d \tau\, \tau
\big(\hat h_3-\hat h_\|^{(t)}\big)\left(\tau\right)
-\frac{2}{\xi^2} \int_\xi^{{\rm sign}(\xi)} \d \tau\, \tau \Big(1-\frac{\xi}{\tau}\Big)
\big(\hat h^{(t)}_\|-\hat\phi_\perp\big)\left(\tau\right).
\end{align}

The relation between the moments may be read off from Eqs.~(\ref{O_v})
-- (\ref{O_s}) as follows:
\begin{align}  
\phi_{\|n}&\equiv\Phi^{(2)}_n,\\
g^{(v)}_{\perp n}&=\Phi^{(3)}_n +  \frac{1}{n+1}
\Big(\Phi^{(2)}_n-\Phi^{(3)}_n\Big),\\
g_{3n}&=\Phi^{(4)}_n -  \frac{1}{n+1}
\Big(\Phi^{(2)}_n-4\Phi^{(3)}_n+3\Phi^{(4)}_n\Big)
+\frac{2}{(n+1)^2}
\Big(\Phi^{(2)}_n-2\Phi^{(3)}_n+\Phi^{(4)}_n\Big),\\
\label{ga_n}
g_{\perp n}^{(a)}&\equiv\Xi^{(3)}_n,\\
\phi_{\perp n}&\equiv\Psi^{(2)}_n,\\
h_{\|n}^{(t)}&=\Psi^{(3)}_n +  \frac{2}{n+2}
\Big(\Psi^{(2)}_n-\Psi^{(3)}_n\Big),\\
h_{3n}&=\Psi^{(4)}_n+  \frac{2}{n+1}
\Big(\Psi^{(2)}_n-\Psi^{(4)}_n\Big)
-\frac{2}{n+2}\Big(\Psi^{(2)}_n-\Psi^{(3)}_n\Big)
-\delta_{n0}\Big(\Psi^{(3)}_n-\Psi^{(4)}_n\Big),\\
h_{\|n}^{(s)}&\equiv\Upsilon^{(3)}_n.
\end{align}
In terms of the moments the relations between old and new wave
functions may be easily inverted; for the wave functions itself
the expression of the new wave functions through the old ones is more
involved. The inverse relations are:
\begin{align}  
\Phi^{(3)}_{n}&=g_{\perp n}^{(v)} +  \frac{1}{n}
\Big(g_{\perp n}^{(v)}-\phi_{\|n}\Big),\qquad n>0\\
\Phi^{(4)}_{n}&=
g_{3n}+  \frac{1}{n-1}
\Big(3g_{3n}-4g_{\perp n}^{(v)}+\phi_{\| n}\Big)
+\frac{1}{n(n-1)}\Big(g_{3n}-4g_{\perp n}^{(v)}+3\phi_{\|n}\Big),\qquad  n>1\\
\Psi^{(3)}_{n}&=h_{\|n}^{(t)}+  \frac{2}{n}
\Big(h_{\|n}^{(t)}-\phi_{\perp n}\Big),\qquad  n>0\\
\Psi^{(4)}_n&=
h_{3n}+  \frac{2}{n-1}
\Big(h_{3n}-h_{\|n}^{(t)}\Big)
-\frac{2}{n(n-1)}\Big(h_{\| n}^{(t)}-\phi_{\perp n}\Big),\qquad  n>1.
\end{align}

The relations between the dynamical and geometrical twist meson wave 
functions show that the dynamical wave functions can be used to determine 
the new geometrical ones and vice versa.
Obviously, the same holds for their moments. In principle, this
allows to determine, e.g., the new wave functions from
the experimental data if these are known for the conventional 
ones. Thus, all of the physical and experimental correlations obtained 
by Ball~{\it et al.}~\cite{ball96,ball98,ball99}, e.g., the asymptotic wave
functions, can be used for the 
geometrical twist wave functions.

\section{Wandzura-Wilczek-type relations}
This section is devoted to the general discussion of geometric Wandzura-Wilczek-type
relations between the conventional wave functions which we have 
obtained by means of the wave functions with a definite geometric twist.
These geometric Wandzura-Wilczek-type relations are independent from the QCD field equations
and the corresponding operator relations. 
The mismatch between the definition of dynamical and geometric twist gives
rise to relations of geometric Wandzura-Wilczek-type which show that the dynamical 
twist functions contain various parts of different geometric twist. 
Consequently, a genuine geometric twist wave function do not give rise
to any geometric Wandzura-Wilczek-type relation. Thereby, we obtain new 
(geometric) Wandzura-Wilczek-type relations and sum rules for the meson wave functions 
of dynamical twist.

We use the notations $\gvh(\xi)=\hat g^{(v),\rm{tw2}}_\perp(\xi)+\hat g^{(v),\rm{tw3}}_\perp(\xi)$, 
where $\hat g^{(v),\rm{tw2}}_\perp(\xi)$  is the genuine twist-2 and 
$\hat g^{(v),\rm{tw3}}_\perp(\xi)$ the genuine
twist-3 part of $\hat g^{(v)}_\perp(\xi)$. 
Substituting (\ref{rel-phi1}) into (\ref{rel-gv}),
we get
\begin{align}
\label{WW-tw2}
\hat g^{(v),\rm{tw2}}_\perp(\xi)&= \int_\xi^{{\rm sign}(\xi)} \frac{\d \tau}{\tau}\,
\hat\phi_\|(\tau),\\
\label{WW-tw3}
\hat g^{(v),\rm{tw3}}_\perp(\xi)&=\gvh(\xi)-\int_\xi^{{\rm sign}(\xi)} \frac{\d \tau}{\tau}\,\hat\phi_\|(\tau),
\end{align}
where (\ref{WW-tw2}) is really the analogue of the Wandzura-Wilczek relation 
for the twist-2 part~\cite{Wandzura} because they argued that the geometric 
twist-3 contribution should be small.
On the other hand, Eq.~(\ref{WW-tw3})
is the Wandzura-Wilczek-type relation of geometric twist-3 and gives the 
information how much is the physical contribution of the geometric twist-3
operator in this non-forward process. 
These relations show that the transverse distribution $\gvh(\xi)$ is related 
to the longitudinal distribution $\hat\phi_\|(\xi)$.
Obviously, this analogy between the Wandzura-Wilczek relations of parton
distribution functions and of meson wave functions
is not surprising because the operator structures are the same 
and the $\rho$-meson polarization vector formally substitutes 
the nucleon spin vector in the Lorentz structures.
The generalization of the geometric Wandzura-Wilczek relation~(\ref{WW-tw2}) for
non-forward distribution amplitudes was discussed in Ref.~\cite{BR00}.
Let me note that the Wandzura-Wilczek relations are obtained 
in Refs.~\cite{ABS94,ball96,ball98}
for $g^{(v),\rm{tw2}}_\perp(u)$ are significantly different (see also~\cite{BL01}). 
The reason is that they used equations of motion operator relations
in order to isolate the dynamical twist-3 part. Therefore, their 
Wandzura-Wilczek relations are dynamical relations 
(dynamical Wandzura-Wilczek-type relations). 
The corresponding geometric Wandzura-Wilczek-type relations for the moments 
are: 
\begin{align}
\label{WW-tw2-n}
g^{(v),\rm{tw2}}_{\perp n}&= \frac{1}{n+1}\,\phi_{\| n},\\
\label{WW-tw3-n}
g^{(v),\rm{tw3}}_{\perp n}&=g^{(v)}_{\perp n}-\frac{1}{n+1}\,\phi_{\| n}, \qquad n>0.
\end{align}
Note that Eq.~(\ref{WW-tw2-n}) was earlier obtained in Ref.~\cite{ball96} 
The reason for this agreement between the twist-2 part of  
the dynamical and geometric Wandzura-Wilczek relation is that the dynamical 
and geometric twist-2 meson wave function coincides. 
Obviously, from the relation (\ref{WW-tw2-n}) for $n=0$ and $n=1$ the 
following sum rules follow:
\begin{align}
\int_{-1}^1\d \xi\,\hat g^{(v)}_{\perp}(\xi) =\int_{-1}^1\d \xi\, \hat\phi_{\|}(\xi)=1,
\qquad
\int_{-1}^1\d \xi\,\xi \hat g^{(v),\rm{tw2}}_{\perp}(\xi) =\frac{1}{2}\int_{-1}^1\d \xi\,\xi \hat\phi_{\|}(\xi).
\end{align}
Using the formulas (\ref{rel-phi1}), (\ref{rel-gv}), (\ref{rel-g3}) and
(\ref{rel-Phi3}), we obtain the integral relations for the function
$\hat g_3(\xi)=\hat g^{\rm tw2}_3(\xi)+\hat g^{\rm tw3}_3(\xi)+\hat g^{\rm tw4}_3(\xi)$ as
\begin{align}
\label{WW-g3-tw2}
\hat g^{\rm tw2}_3(\xi)=&-\int_\xi^{{\rm sign}(\xi)}\frac{\d \tau}{\tau}            
\Big\{\hat\phi_\|(\tau)+2\ln\Big(\frac{\xi}{\tau}\Big)\hat\phi_\|(\tau)\Big\},\\
\label{WW-g3-tw3}
\hat g^{\rm tw3}_3(\xi)=&\,4\int_\xi^{{\rm sign}(\xi)} \frac{\d \tau}{\tau}\,\gvh(\tau)
              +4\int_\xi^{{\rm sign}(\xi)}\frac{\d \tau}{\tau^2}\int_\tau^{{\rm sign}(\tau)}\d\omega
              \big(\gvh-\hat\phi_\|\big)(\omega)\nonumber\\
             &+4\int_\xi^{{\rm sign}(\xi)} \frac{\d \tau}{\tau}\ln\Big(\frac{\xi}{\tau}\Big)\gvh(\tau)
              +4\int_\xi^{{\rm sign}(\xi)} \frac{\d \tau}{\tau^2}\ln\Big(\frac{\xi}{\tau}\Big)
              \int_\tau^{{\rm sign}(\tau)}\d \omega\big(\gvh-\hat\phi_\|\big)(\omega),\\
\label{WW-g3-tw4}
\hat g^{\rm tw4}_3(\xi)=&\,\hat g_3(\xi)-\int_\xi^{{\rm sign}(\xi)} \frac{\d \tau}{\tau}
		\big(4\gvh-\hat\phi_\|\big)(\tau)
              -4\int_\xi^{{\rm sign}(\xi)}\frac{\d \tau}{\tau^2}\int_\tau^{{\rm sign}(v)}\d \omega
              \big(\gvh-\hat\phi_\|\big)(\omega)\nonumber\\
             &-2\int_\xi^{{\rm sign}(\xi)} \frac{\d \tau}{\tau}\ln\Big(\frac{\xi}{\tau}\Big)
		\big(2\gvh-\hat\phi_\|\big)(\tau)
              -4\int_\xi^{{\rm sign}(\xi)} \frac{\d \tau}{\tau^2}\ln\Big(\frac{\xi}{\tau}\Big)
              \int_\tau^{{\rm sign}(\tau)}\d \omega\big(\gvh-\hat\phi_\|\big)(\omega).
\end{align}
Due to the fact that $\hat g_3(\xi)$ contains twist-2, twist-3 as well as twist-4,
we have obtained three integral relations. For example, Eq.~(\ref{WW-g3-tw2})
demonstrates that the twist-2 part $\hat g^{\rm tw2}_3(\xi)$ can be expressed in terms
of the twist-2 function $\hat\phi_\|(\xi)$. Additionally, the twist-3 part 
$\hat g^{\rm tw3}_3(\xi)$ 
is given in Eq.~(\ref{WW-g3-tw3}) in terms of the functions 
$\hat\phi_\|(\xi)$ and $\gvh(\xi)$.
Eq.~(\ref{WW-g3-tw4}) expresses how can
be obtained the twist-4 part $\hat g^{\rm tw4}_3(\xi)$ from $\hat g_3(\xi)$.
For the wave function moments we obtain:
\begin{align}
\label{WW-g3-tw2-n}
g^{\rm tw2}_{3 n}=&-\frac{1}{n+1}\,\phi_{\| n}+\frac{2}{(n+1)^2}\,\phi_{\| n},\\
\label{WW-g3-tw3-n}
g^{\rm tw3}_{3 n}=&\,\frac{4}{n+1}\, \gvn
              +\frac{4}{(n+1)n}\Big(\gvn-\phi_{\| n}\Big)
              -\frac{4}{(n+1)^2}\,\gvn
              -\frac{4}{(n+1)^2 n}\Big(\gvn-\phi_{\| n}\Big),\qquad n>0\\
g^{\rm tw4}_{3 n}=&\,g_{3 n}
              -\frac{1}{n+1}\Big(4\gvn-\phi_{\| n}\Big)
              -\frac{4}{(n+1)n}\Big(\gvn-\phi_{\| n}\Big)
              +\frac{2}{(n+1)^2}\Big(2\gvn-\phi_{\| n}\Big)\nonumber\\
              &\qquad +\frac{4}{(n+1)^2 n}\Big(\gvn-\phi_{\| n}\Big),\qquad n>1.
\end{align}
For $n=0,1$ in (\ref{WW-g3-tw2-n}) we find the sum rules
\begin{align}
\int_{-1}^1\d \xi\, \hat g_3(\xi) =  \int_{-1}^1 \d \xi\,\hat \phi_{\|}(\xi)=1,
\qquad
\int_{-1}^1\d \xi\, \xi \hat g^{\rm tw2}_3(\xi)=0,
\end{align}
as well as for $n=1$ in (\ref{WW-g3-tw3-n})
\begin{align}
\int_{-1}^1\d \xi\, \xi \hat g^{\rm tw3}_3(\xi) = \int_{-1}^1 \d \xi\,\xi\big(
                                         2\hat g_\perp^{(v)}-\hat\phi_{\|}\big)(\xi).
\end{align}

Let us now discuss the geometric Wandzura-Wilczek-type relations for the
chiral-odd meson wave functions.
If we substitute (\ref{rel-phi2}) into (\ref{rel-ht}) and using
$\hat\htt(\xi)=\hat h^{(t),\rm{tw2}}_\|(\xi)+\hat h^{(t),\rm{tw3}}_\|(\xi)$
we derive
\begin{align}
\label{WW-JJ-tw2}
\hat h^{(t),\rm{tw2}}_\|(\xi)&= 2\xi\int_\xi^{{\rm sign}(\xi)} \frac{\d \tau}{\tau^2}\,\hat \phi_\perp(\tau),\\
\label{WW-JJ-tw3}
\hat h^{(t),\rm{tw3}}_\|(\xi)&=\hat h^{(t)}_\|(\xi)- 2\xi\int_\xi^{{\rm sign}(\xi)} 
\frac{\d\tau}{\tau^2}\,\hat \phi_\perp(\tau),
\end{align}
where (\ref{WW-JJ-tw2}) is the analogues of the twist-2 Wandzura-Wilczek-type relation of 
the nucleon structure function $h_L(z)$ which was obtained by Jaffe and Ji ~\cite{Jaf92}. 
These Wandzura Wilczek-type relations  point out that the longitudinal distribution 
$\hat \htt(\xi)$ is related 
to the transverse distribution $\hat\phi_\perp(\xi)$.
Dynamical Wandzura-Wilczek relations were obtained in Ref.~\cite{ball98}. 
Obviously, (\ref{WW-JJ-tw3}) is the corresponding twist-3 relation. 
The relations for the moments read:
\begin{align}
\label{WW-JJ-tw2-n}
h^{(t),\rm{tw2}}_{\| n}&=  \frac{2}{n+2}\,\phi_{\perp n},\\
\label{WW-JJ-tw3-n}
h^{(t),\rm{tw3}}_{\| n}&=h^{(t)}_{\| n}-\frac{2}{n+2}\,\phi_{\perp n},\qquad n>0.
\end{align}
For $n=0,1$ in (\ref{WW-JJ-tw2-n}) we observe the following sum rules
\begin{align}
\int_{-1}^1\d \xi\, \htth(\xi) =  \int_{-1}^1\d \xi\,\hat \phi_{\perp}(\xi)=1,
\qquad
\int_{-1}^1\d \xi\, \xi \hat h^{(t),\rm tw2}_{\|}(\xi) = \frac{2}{3} \int_{-1}^1\d \xi\, \xi \hat\phi_\perp(\xi).
\end{align}

Now, using the formulas (\ref{rel-phi2}), (\ref{rel-ht}), (\ref{rel-h3}) and
(\ref{rel-Psi3}), we obtain the integral relations for the function
with $\hat h_3(\xi)=\hat h^{\rm tw2}_3(\xi)+\hat h^{\rm tw3}_3(\xi)+\hat h^{\rm tw4}_3(\xi)$ as
\begin{align}
\label{WW-h3-tw2}
\hat h^{\rm{tw2}}_3(\xi)&=2\int_\xi^{{\rm sign}(\xi)}\frac{\d \tau}{\tau}\,\hat\phi_\perp(\tau)
                     -2\xi\int_\xi^{{\rm sign}(\xi)}\frac{\d \tau}{\tau^2}\,\hat\phi_\perp(\tau), \\
\label{WW-h3-tw3}
\hat h^{\rm tw3}_3(\xi)&=2\xi\int_\xi^{{\rm sign}(\xi)} \frac{\d \tau}{\tau^2}\,\htth (\tau)
              +4\xi\int_\xi^{{\rm sign}(\xi)}\frac{\d \tau}{\tau^3}\int_\tau^{{\rm sign}(\tau)}\d \omega                                \big(\htth-\hat\phi_\perp\big)(\omega),\\
\label{WW-h3-tw4}
\hat h^{\rm tw4}_3(\xi)&=\hat h_3(\xi)-2\int_\xi^{{\rm sign}(\xi)}\frac{\d \tau}{\tau}\,\hat\phi_\perp(\tau)
               -2\xi\int_\xi^{{\rm sign}(\xi)} \frac{\d \tau}{\tau^2}\big(\htth-\hat\phi_\perp\big) (\tau)
              -4\xi\int_\xi^{{\rm sign}(\xi)}\frac{\d \tau}{\tau^3}\int_\tau^{{\rm sign}(\tau)}\d \omega                                \big(\htth-\hat\phi_\perp\big)(\omega)
\end{align}
and for the moments
\begin{align}
\label{WW-h3-tw2-n}
h^{\rm{tw2}}_{3 n}&=\frac{2}{n+1}\,\phi_{\perp n}
                     -\frac{2}{n+2}\,\phi_{\perp n}, \\
\label{WW-h3-tw3-n}
h^{\rm tw3}_{3 n}&=\frac{2}{n+2}\,\httn
              +\frac{4}{(n+2)n}\Big(\httn-\phi_{\perp n}\Big),\qquad n>0\\
h^{\rm tw4}_{3n}&=h_{3n}-\frac{2}{n+1}\,\phi_{\perp n}
               -\frac{2}{n+2}\Big(\httn-\phi_{\perp n}\Big) 
              -\frac{4}{(n+2)n}\Big(\httn-\phi_{\perp n}\Big),\qquad n>1.
\end{align}
Thus, Eq.~(\ref{WW-h3-tw2})
means that the twist-2 part $\hat h^{\rm tw2}_3(\xi)$ can be expressed in terms
of the twist-2 function $\hat\phi_\perp(\xi)$.
Additionally, the twist-3 part $\hat h^{\rm tw3}_3(\xi)$ 
is given in Eq.~(\ref{WW-h3-tw3}) in terms of the functions $\hat\phi_\perp(\xi)$ 
and $\htth(\xi)$.
Eq.~(\ref{WW-h3-tw4}) gives the subtraction rule in order to obtain the 
twist-4 part $\hat h^{\rm tw4}_3(\xi)$ from the original function $\hat h_3(\xi)$. 
Obviously, the same is genuine for the corresponding wave function moments.

For $n=0,1$ in (\ref{WW-h3-tw2-n}) we observe the following sum rules
\begin{align}
\int_{-1}^1\d \xi\,\hat h_3(\xi) = \int_{-1}^1\d \xi\, \hat\phi_{\perp}(\xi)=1,
\qquad
\int_{-1}^1\d \xi\, \xi \hat h^{\rm tw2}_{3}(\xi) = \frac{1}{3} \int_{-1}^1\d \xi\, \xi \hat\phi_\perp(\xi),
\end{align}
and for $n=1$ in (\ref{WW-h3-tw3-n})
\begin{align}
\int_{-1}^1\d \xi\, \xi\hat h^{\rm tw3}_{3}(\xi) = \frac{1}{3} \int_{-1}^1\d \xi\, \xi \big( 6\htth(\xi)-4\hat\phi_\perp(\xi)\big).
\end{align}

Because the functions $\gah(\xi)$ and $\hsh(\xi)$ are pure geometric as well as dynamical 
twist-3 functions, there is no mismatch between the dynamical and geometric twist
and no geometric Wandzura-Wilczek-type relation occurs. 
However, dynamical Wandzura-Wilczek relations for $\gah(\xi)$ and $\hsh(\xi)$
were obtained by Ball {\it et al.}~\cite{ball96,ball98}
by using the QCD equations of motion. Here I am discussing the relation for
$g^{(a)}_{\perp n}$. Neglecting quark masses and three-particle quark-antiquark-gluon 
operators of twist-3 one recovers the operator relation~(see~\cite{BB88})
\begin{align}
\label{total1}
\big[\bar{u}(\lcx)\gamma_\alpha\gamma_5 d(-\lcx)\big]^{\rm tw3}=
\ii\, \epsilon_\alpha^{\ \,\beta\mu\nu}\lcx_\beta \int_0^1\d u\, u\, \hat\pd_\mu
\big[\bar{u}(u\lcx)\gamma_\nu d(-u\lcx)\big]+\ldots
\end{align}
where $\hat\pd_\mu$ is the so-called total derivative which translates the expansion point 
$y$:
\begin{align}
\label{tot-der}
\hat\pd_\mu
\big[\bar{u}(u\lcx)\gamma_\nu d(-u\lcx)\big]\equiv
\frac{\pd}{\pd y^\mu}
\big[\bar{u}(y+u\lcx)\gamma_\nu d(y-u\lcx)\big]\big|_{y\rightarrow 0}.
\end{align}
The total derivative on the r.h.s. of Eq.~(\ref{total1}) is calculated as
\begin{align}
\label{total2}
\hat\pd_\mu
\big[\bar{u}(u\lcx)\gamma_\nu d(-u\lcx)\big]=
\bar{u}(u\lcx)\big(\LD D\!_\mu+\RD D\!_\mu\big)\gamma_\nu d(-u\lcx)
-\ii\int_{-u}^{u}\d v\, \bar{u}(u\lcx)\lcx^\rho F_{\rho\mu}(v\lcx)\gamma_\nu d(-u\lcx),
\end{align}
where $F_{\mu\nu}$ is the gluon field strength.
Obviously, the second operator on the r.h.s. of Eq.~(\ref{total2}) is a 
Shuryak-Vainshtein type operator having minimal twist-3. The other operator
on the r.h.s. of Eq.~(\ref{total2}) has, in general, twist-2, 3 and 4. 
But, after multiplying by the Levi-Civita tensor $\epsilon_{\alpha\beta\mu\nu}$
only the twist-3 part survives and gives contributions which are not small. 
Therefore, the operator identity (\ref{total1}) is a genuine twist-3 relation. 
The twist-3 matrix element of the first operator on the r.h.s. of Eq.~(\ref{total2}) 
is given by
\begin{align}
\label{O_total}
\ii\, \epsilon_\alpha^{\ \,\beta\mu\nu}\lcx_\beta
&\langle 0|
\bar{u}(\lcx)\big(\LD D\!_\mu+\RD D\!_\mu\big)\gamma_\nu d(-\lcx)
|\rho(P,\lambda)\rangle
= f_\rho m_\rho
\epsilon_\alpha^{\ \,\beta\mu\nu}
e^{(\lambda)}_{\perp\beta} p_\mu\lcx_\nu
\int_{-1}^1\d \xi\, \hat\Omega^{(3)}(\xi) e_0(\ii\zeta \xi),
\end{align}
where $\hat\Omega^{(3)}(\xi)$  is the corresponding meson wave function 
of twist-3.

Taking the matrix element of the operators on both sides of Eq.~(\ref{total1}), 
using Eqs.~(\ref{O_5v}), (\ref{rel-ga}), and (\ref{O_total}),
and neglecting quark masses and trilocal operators, one obtains a geometric 
relation between the meson wave functions $\hat g^{(a)}_\perp(\xi)$ and $\hat\Omega^{(3)}(\xi)$
imposed by the QCD equations of motion as 
\begin{align}
\label{dyn-rel1}
\hat g^{(a)}_\perp(\xi)=2\xi \int_\xi^{\text{sign}(\xi)}\frac{\d\tau}{\tau^2}\, \hat\Omega^{(3)}(\tau),
\end{align}
and for the meson wave function moments 
\begin{align}
\label{dyn-rel2}
g^{(a)}_{\perp n}=\frac{2}{n+2}\, \Omega^{(3)}_{ n},
\end{align}
where $\Omega^{(3)}_{ n}$ are the geometric twist-3 wave function moments.
This relation (\ref{dyn-rel2}) reproduces the fact that the moments 
$g^{(a)}_{\perp n}$ have genuine geometric twist-3 (see Eq.~(\ref{ga_n}))
and no geometric twist-2 contributions from the total derivative come into the game. 

But this is not what Ball {\it et al.}~\cite{ball96,ball98} did.
They have calculated the r.h.s. of Eq.~(\ref{total1}) by means of
\begin{align}
\hat\pd_\mu
\langle 0|\bar{u}(\lcx)\gamma_\nu d(-\lcx)
|\rho(P,\lambda)\rangle
\end{align}
and Eq.~(\ref{eq:vda}).
Eventually, they obtained the dynamical Wandzura-Wilczek-type relation~\cite{ball96}
\begin{align}
\label{dyn-rel3}
\hat g^{(a)}_\perp(\xi)=2\xi \int_\xi^{\text{sign}(\xi)}\frac{\d\tau}{\tau^2}\, \hat g^{(v)}_\perp(\tau),
\end{align}
and for the  moments 
\begin{align}
\label{dyn-rel4}
g^{(a)}_{\perp n}=\frac{2}{n+2}\, g^{(v)}_{\perp n}.
\end{align}
Obviously, Eqs.~(\ref{dyn-rel3}) and (\ref{dyn-rel4}) are relations between
equal dynamical twist. But, the l.h.s. of these relations has geometric twist-3
and the r.h.s. has geometric twist-2 and twist-3. Therefore, a dynamical 
Wandzura-Wilczek relation is a relation for equal dynamical twist
in contrast to geometric
Wandzura-Wilczek relations which are relations for equal geometric twist.
Note that the Wandzura-Wilczek-like relations used in Refs.\cite{gauge}
are dynamical ones.

Similar arguments are valid for all QCD equations of motion operator relations
used in Ref.~\cite{ball98}.
Thus, from the group theoretical point of view, it is a bit misleading to 
claim that an additional geometric twist-2 contribution is produced by 
geometric twist-3 operators with a total derivative
and that the QCD equations of motion give a relation 
between distribution functions of different geometric twist
in exclusive processes and off-forward scattering. 
From this point of view, the geometric twist is a proper concept for 
classifying non-forward inclusive or general exclusive matrix-elements.


\section{Conclusions}
Extending an earlier study~\cite{L00}, we discussed the model-independent 
classification of meson wave functions with respect to geometric twist. 
We gave the relations between 
the geometric twist and Ball and Braun's dynamical twist wave functions. 
These relations demonstrate the interrelations between the different twist definitions.
Consequently, these relations are ``transformation rules'' between the dynamical 
and geometric twist wave functions.

The main results of this letter are geometric Wandzura-Wilczek-type relations
between the dynamical twist distributions which we have obtained by means of 
our ``transformation rules''. 
The reason of these geometric Wandzura-Wilczek-type relations is the mismatch 
of the dynamical twist with respect to geometric twist. 
Because we have not used the QCD equations of motion, 
the geometric Wandzura-Wilczek-type relations are based on a no-dynamical level 
and are model-independent. 
Additionally, I have discussed the difference between dynamical and geometric
Wandzura-Wilczek-type relations.
 

\acknowledgments
The author is grateful to P.~Ball, J.~Eilers, B.~Geyer, S.~Neumeier, 
D.~Robaschik and C.~Weiss for stimulating discussions.
He acknowledges the Graduate College
``Quantum field theory'' at Centre for Theoretical Sciences of
Leipzig University for financial support and the CERN/TH-division 
for the support during his stay in Geneva.

\end{document}